\documentclass{cpbtex}
\usepackage{makecell}
\usepackage{eqparbox}
\usepackage{graphicx}
\usepackage{subfig}
\usepackage{enumerate}
\usepackage{lmodern}
\usepackage[T1]{fontenc}

\begin{document}

\title{SolarDesign: An online photovoltaic device simulation and design platform\thanks{
Project supported by the Scientific Research Project of China Three Gorges Corporation (Grant No. 202203092).
}}


\author{Wei E. I. Sha $^{1}$\thanks{Corresponding author. E-mail:~weisha@zju.edu.cn}, 
\ Xiaoyu Wang$^{2}$, \ Wenchao Chen$^{3,1}$, \ Yuhao Fu$^{4}$, \ Lijun Zhang$^{2}$, \ Liang Tian$^{3,1}$, \\ Minshen Lin$^{5}$, \ Shudi Jiao$^{1}$, \ Ting Xu$^{1}$, \ Tiange Sun$^{6}$, \ Dongxue Liu$^{6}$\thanks{Corresponding author. E-mail:~liu$\_$dongxue@ctg.com.cn}\\
$^{1}${College of Information Science and Electronic Engineering,}\\{Zhejiang University, Hangzhou 310027, China}\\  
$^{2}${State Key Laboratory of Integrated Optoelectronics,}\\{Key Laboratory of Automobile Materials of MOE,}\\ {Key Laboratory of Material Simulation Methods \& Software of MOE,}\\ {School of Materials Science and Engineering, Jilin University, Changchun 130012, China}\\ 
$^{3}${ZJU-UIUC Institute, International Campus, Zhejiang University, Haining 314400, China}\\
$^{4}${Key Laboratory of Superhard Materials, College of Physics,}\\ {Jilin University, Changchun 130012, China}\\
$^{5}${College of Electrical Engineering, Zhejiang University, Hangzhou 310027, China}\\
$^{6}${Science and Technology Research Institute,}\\{China Three Gorges Corporation, Beijing 101199, China}\\
}  




\date{\today}
\maketitle

\begin{abstract}
SolarDesign (https://solardesign.cn/) is an online photovoltaic device simulation and design platform that provides engineering modeling analysis for crystalline silicon solar cells, as well as emerging high-efficiency solar cells such as organic, perovskite, and tandem cells. The platform offers user-updatable libraries of basic photovoltaic materials and devices, device-level multi-physics simulations involving optical-electrical-thermal interactions, and circuit-level compact model simulations based on detailed balance theory. Employing internationally advanced numerical methods, the platform accurately, rapidly, and efficiently solves optical absorption, electrical transport, and compact circuit models. It achieves multi-level photovoltaic simulation technology from ``materials to devices to circuits'' with fully independent intellectual property rights. Compared to commercial software, the platform achieves high accuracy and improves speed by more than an order of magnitude. Additionally, it can simulate unique electrical transport processes in emerging solar cells, such as quantum tunneling, exciton dissociation, and ion migration.
\end{abstract}



\section{Introduction}
Solar cells directly convert solar energy into electrical energy, playing a crucial role in addressing the energy crisis and adjusting China's energy structure for sustainable development. Currently, the share of photovoltaic power in China's energy mix is relatively low, with silicon-based cells dominating the market. However, further reducing the cost of silicon cells while enhancing their photoelectric conversion efficiency has encountered significant challenges. Therefore, the development of emerging solar cells based on new materials and structures, such as organic, perovskite, and tandem cells, is strategically important for improving energy conversion efficiency, reducing industrial costs, and advancing photovoltaic technology.

Solar cells exhibit diverse materials and device structures, complex surface/interface charge interactions, and varying carrier transport mechanisms. For instance, in silicon cells, carrier quantum tunneling~\cite{1}; in organic cells, exciton diffusion and dissociation~\cite{2}; and in perovskite cells, ion migration and accumulation~\cite{3,4}. These differing charge transport processes introduce new scientific challenges and issues regarding device efficiency, stability, and measurement standards. Conducting ``optical-electrical-thermal'' multi-physics simulations to develop universal and specific theoretical models, together with accurate and fast simulation methods is crucial. The simulations (involving models and methods) aid in understanding device physics, optimizing designs, establishing standard testing platforms, and strengthening the photovoltaic industry by providing solid theoretical foundations and reliable simulation tools.

Existing commercial software such as COMSOL~\cite{5}, Silvaco~\cite{6}, and Lumerical FDTD~\cite{7} can only model traditional silicon and III-V material photovoltaic devices. They lack specialized models for emerging solar cell technologies, and they suffer from low computational efficiency, high price, and expensive annual updates. Online photovoltaic educational platforms like PVLighthouse~\cite{8} and PVEducation~\cite{9} offer limited coverage of the working principles, theoretical models, and simulation methods for emerging photovoltaic devices. With China's photovoltaic industry approaching a trillion-dollar valuation, there is an urgent need to develop a simulation design platform with completely independent intellectual property. Establishing such an online platform with a cloud service architecture, and making it accessible to universities, research institutes, and companies, is of significant importance for strengthening foundational research in photovoltaics and enhancing the quality of photovoltaic talent development.

\section{Function}

The SolarDesign platform (https://solardesign.cn/) is an online simulation platform developed collaboratively by China Three Gorges Corporation, Zhejiang University, and Jilin University to serve the photovoltaic research and industry. It is suitable for engineering modeling and analysis of various photovoltaic devices, including traditional silicon cells, as well as new high-efficiency solar cells such as organic, perovskite, and tandem cells. The platform features: (1) updatable database including cloud libraries of fundamental photovoltaic materials and devices; (2) multi-physics simulation including ``optical-electrical-thermal'' device-level modeling; (3) compact model simulation including circuit-level modeling based on detailed balance theory.

Specific functions are described in detail: (1) optical simulation including optical power absorptance and reflectance, exciton/carrier generation rate, planar and textured surface designs, and incident light spectrum matching (Solar AM 1.5G and user-defined spectra); (2) general electrical response simulation including current density-voltage (J-V) curve, characteristic parameters, electric potential, carrier concentrations/current densities, energy band diagram, and non-radiative recombination; (3) specialized electrical response simulation including exciton diffusion and dissociation for organic cells, ion migration and accumulation for perovskite cells, as well as hot electron emission and quantum tunneling for tandem cells and tunnel oxide passivated contact cells; (4) circuit simulation including diode equation based traditional model (i.e. short-circuit current, dark current, ideality factor, series and shunt resistances five-parameter model), detailed balance theory based new model (i.e. bulk/surface/Auger nonradiative dark currents, series and shunt resistances five-parameter model), and device efficiency bottleneck analyses (i.e. quantitative analyses of optical loss, three types of non-radiative recombination losses, and Ohmic losses); (5) material library covering silicon, III-V/II-VI materials, organic, and perovskite halides with material parameters of bandgap, complex refractive index, optical absorption coefficient, optical transition intensity, exciton binding energy, carrier effective mass, dielectric constant, electron affinity, and metal work function; (6) device simulation library covering various types of solar cells; (7) additional features such as cloud-based material and device library processing, distributed multi-user concurrent computing, dynamic display of simulation results, intelligent graphical user interface, advanced technical forum, and video tutorials and help.

To show the advantages of SolarDesign, Table 1 compares the functions of different photovoltaic simulation platforms. 

\begin{table}[h!]
  \begin{center}
    \caption{Functions of different photovoltaic simulation platforms}
    \begin{tabular}{c|c|c|c|c|c} 
    \hline
      {Name} & {Exciton Dissociation} & {Ion Migration} & {Tunneling} & {Bottleneck Analyses} & {Libraries}\\
      \hline
      Quokka3 & No & No & Yes & No & No \\
      PC1D & No & No & No & No & No \\
      SCAPS & No & No & Yes & No & No \\
      IonMonger & No & Yes & No & No & No \\
      DriftFusion & No & Yes & No & No & No \\
      SolarDesign & Yes & Yes & Yes & Yes & Yes\\
      \hline
    \end{tabular}
  \end{center}
\end{table}

\begin{table}[h!]
  \begin{center}
    \caption{Cross-scale photovoltaic simulation at three levels}
    \begin{tabular}{c|c|c|c|c} 
    \hline
      {Levels} & {Scales} & {Equations} & {Algorithms} & {Interface Parameters}\\
      \hline
      circuit & $\sim$ 1 $\mathrm{\mu m}$  & \makecell{detailed balance theory\\ diode equations\\Kirchhoff's laws} & \makecell{modified nodal analysis\\ Newton’s method} & \makecell{series resistance\\shunt resistance\\ideality factor\\dark currents} \\
      \hline
       device & $\sim$ 10 $\mathrm{nm}$  & \makecell{Maxwell's equations\\drift-diffusion model\\specific electrical models} & \makecell{finite difference\\ finite element\\finite volume} & \makecell{J-V curve\\key parameters} \\
      \hline
      material & $\sim$ 1 $\mathrm{nm}$  & \makecell{density-functional\\theory} & \makecell{plane-wave pseudopotentials\\projected augmented-\\wave pseudopotentials} & \makecell{dielectric constant\\refractive index\\electron affinity\\bandgap\\effective mass} \\
      \hline
    \end{tabular}
  \end{center}
\end{table}

\section{Method}
The platform has achieved cross-scale photovoltaic simulation analyses at three levels: materials, devices, and circuits. Table 2 lists the scale ranges, theoretical equations, basic algorithms, and interface parameters between different levels of simulation. More technical details can be found in the following parts: 

\subsection{Material Simulation}
    
To calculate physical properties of photovoltaic materials, first-principles calculations are performed within the framework of density-functional theory (DFT)~\cite{20} using the plane-wave pseudopotential method as implemented in the Vienna Ab initio Simulation Package~\cite{21,22}. The electron-ion interactions were described by using the projected augmented wave pseudopotentials~\cite{23}. We adopted the generalized gradient approximation formulated by Perdew, Burke, and Ernzerhof as exchange-correlation functional~\cite{24}. The Heyd-Scuseria-Ernzerhof functional was then used for the bandgap correction~\cite{25,26}. Structure optimization was achieved using the conjugate gradient technique until energies converged to $10^{-4}$ eV. For organic semiconductor materials, we constructed structurally ordered molecular crystals, taking periodic boundary conditions into account during the calculations. The hybrid perovskite materials with specific alloy compositions were simulated by constructing supercells, and for the chemical compositions that cannot be directly simulated, we used Vegard's law to fit the properties. To properly take into account the long-range van der Waals (vdW) interaction that is non-negligible for hybrid perovskites involving organic molecules, the optB86b-vdW functional was adopted~\cite{27}. The process of DFT-based calculations and analysis of the calculated results were carried out using our in-house developed codes, the Jilin artificial intelligence-aided material design integrated package~\cite{28,29}.

The calculated physical properties include: bandgap, work function, electron affinity, dielectric constant, electronic band structure, density of states, effective mass, exciton binding energy, optical absorption spectrum, complex refractive index, and optical transition intensity. Most of these properties can be obtained through electronic structure analysis. Optical properties can be derived from the frequency-dependent dielectric function~\cite{30}, exciton binding energy is calculated using the hydrogenic model~\cite{31}, and optical transition intensity is proportional to the square of the transition dipole moments. 

\subsection{Optical Simulation}
To calculate optical (absorption) properties of photovoltaic devices, mode-matching methods~\cite{10} are applied. Electromagnetic fields at each layer of the multi-layer device structure are expanded as plane waves, and Maxwell's equations are solved using the continuity conditions for the tangential components of the electric and magnetic fields. Furthermore, the generation rate~\cite{11} is calculated based on the electric field distribution in the active layer, the corresponding complex refractive index, and the incident light spectrum. Additionally, for textured surfaces, we use the $4n^2$ (Yablonovitch) absorption limit~\cite{12,13} and set the texture factor ranging from 0 to 1, with 1 indicating the $4n^2$ absorption limit.

\subsection{Electrical Simulation}
For general models, non-uniform spatial discretization (Scharfetter-Gummel scheme), semi-implicit time stepping, and Gummel iteration~\cite{14, 15} are employed to solve the drift-diffusion model, which includes the Poisson's equation and the current continuity equations for electrons and holes.

For specialized models, different methods should be adopted for different types of solar cells. 
Regarding quantum tunneling, we employ the non-local band-to-band tunneling models and WKB (Wentzel-Kramers-Brillouin) approximation, seamlessly integrated with the drift-diffusion model~\cite{1}. Regarding exciton dynamics, we use the Langevin bimolecular recombination model and a mixed dissociation model, which combines local diffusion-dissociation (Onsager-Braun model) with direct delocalization~\cite{2}. Regarding ion dynamics, we apply the ion migration and accumulation model, involving continuity equations for positive and negative ion flows (similar to carrier current continuity equations, including drift and diffusion terms, but excluding generation and recombination terms), initial concentration conditions, and annihilation boundary conditions~\cite{3}.

\subsection{Circuit Simulation}

Based on detailed balance theory, we compute radiative recombination using blackbody radiation spectrum and decompose non-radiative recombination into bulk/surface Shockley-Read-Hall (SRH) recombination and Auger recombination. Meanwhile, Ohmic loss from electrodes, low-mobility transport layers, and interfaces between the transport and active layers is represented by series resistance; Ohmic loss from shunt current caused by defects and pinholes is represented by shunt resistance. All of the above lead to the development of a new circuit model for intelligent quantification of solar cell efficiency losses~\cite{16,17,18}.

Additionally, the modified nodal analysis (MNA) and high-dimensional Newton's method are used to solve circuit models for both single solar cell and large-scale module configurations.

\section{Input and Output}

\subsection{Optical Simulation}
The input parameters include device structure (thickness of each layer), incident light spectrum, and complex refractive indices of all materials. Among them, the complex refractive index can be obtained from the Material Library, and the light spectrum can be the Solar AM 1.5G spectrum or user-defined spectrum.

The output results include optical absorptance and reflectance, carrier (exciton) generation rate, and maximum short-circuit current.
    
\subsection{General Electrical Simulation}
The input parameters of the non-electrode layers include relative dielectric constant, electron affinity, bandgap, electron/hole mobility, minority carrier lifetime of bulk/surface SRH recombination, Auger recombination coefficient, effective density of states in the conduction/valence band, and donor/acceptor doping concentration. Moreover, users can choose to import the generation rate from the optical output results or input the averaged generation rate themselves. The input parameters of the electrode layers include work function and contact type (Schottky/Ohmic). To simulate the transient current density-voltage curve, it is necessary to provide the voltage scanning range. To simulate the steady-state operating point, a given operating voltage is required. Among them, the relative dielectric constant, electron affinity, bandgap, and effective density of states can be obtained from the Material Library.

The output results of transient simulation include current density-voltage curve and characteristic parameters (short-circuit current, open-circuit voltage, fill factor, power conversion efficiency). The output results of steady-state simulation include electron/hole current, electron/hole concentration, electric potential, non-radiative recombination, and energy band diagram.

\subsection{Specialized Electrical Simulation}
The additional input parameters for perovskite materials include initial concentration of positive/negative ions, mobility of positive/negative ions, and voltage scanning rate. The transient output results include current density-voltage hysteresis curves (forward-reverse scan) and corresponding characteristic parameters. The steady-state output results include the net ion concentration distribution in addition to the general output.

The additional input parameters for organic materials include the delocalization ratio of excitons, exciton lifetime, exciton mobility, exciton radius, and bimolecular recombination coefficient. The steady-state output results include the exciton concentration distribution and exciton dissociation probability in addition to the general output.

The additional input parameters for tunneling layers include the electron/hole tunneling coefficient (relative effective mass).

\subsection{Circuit Simulation}
Whether for traditional or new circuit models, the input parameters are the experimentally measured current density-voltage curves, and the output results are the numerically fitted current density-voltage curves and fitting errors (root mean square error and mean absolute error). Furthermore, the output results of traditional circuit model include five parameters of the model (short-circuit current, saturated dark current, ideality factor, series/shunt resistances). The output results of the new circuit model also include five parameters of the model (saturated dark currents of bulk SRH, surface SRH, and Auger recombination, series/shunt resistances) and the quantified proportion of device efficiency loss (bulk SRH recombination, surface SRH recombination, Auger recombination, series/shunt resistances).

\section{Performance}

The simulation results for silicon, organic, and perovskite solar cells obtained from the online platform will be presented and compared with those from commercial software. Optical and electrical parameters for all test cases are available in the Public Device Library of the online platform.

\subsection{Silicon solar cell}
Device structure of a tunnel oxide passivated contact (TOPCon) silicon solar cell is set as: Ag/SiNx [70 nm]-c-Si ($\rm{p}^+$) [300 nm]-c-Si (n) [200 $\rm \mu$m]-SiO2 [1 nm]-c-Si ($\rm{n}^{++}$) [30 nm]-Ag [1000 nm]. Compared to Silvaco, the calculation error of current density-voltage (J-V) characteristics by SolarDesign is lower than 1 \% (See Fig. 1). Particularly, the computer time and memory cost are reduced by more than 10 times. Moreover, the short-circuit current (Jsc), open-circuit voltage (Voc), fill factor (FF), and power conversion efficiency (PCE) of the TOPCon cell are $42.69\,\,\mathrm{mA/cm^2}$, 0.71 V, 0.85, and 25.64\%, respectively. The Shockley–Queisser limit with 33 \% efficiency could be approached by both near-perfect Sunlight absorption above the bandgap of silicon and ignorable non-radiative recombination in the device. Additionally, the continuity of classical total current (that is the sum of the drift and diffusion currents) can be satisfied at all device regions except for the tunnel oxide layer, where quantum tunneling occurs (See Fig. 2).

\begin{center}
    \includegraphics[width=0.7\linewidth]{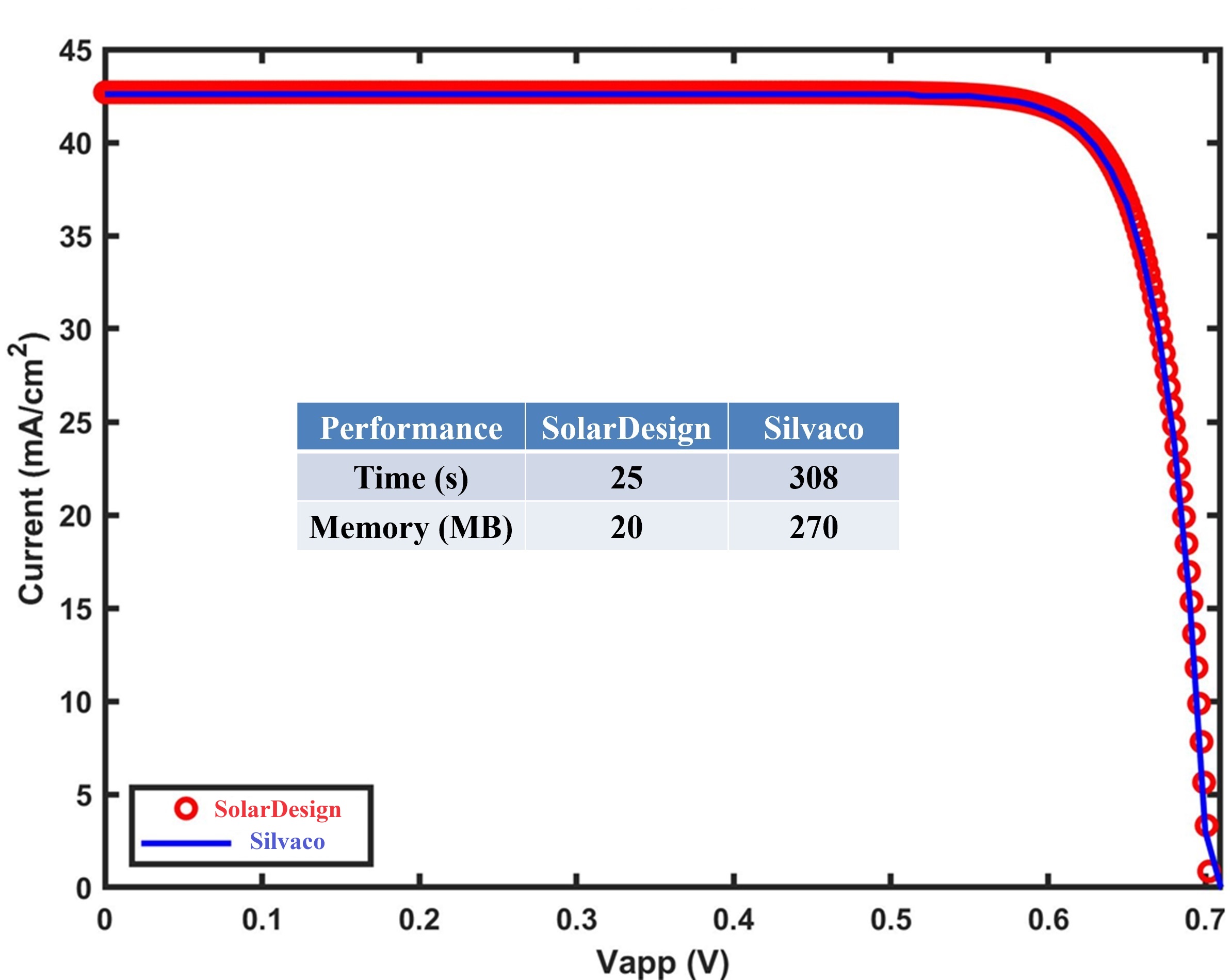}\\[5pt]  
    \parbox[c]{15.0cm}{\footnotesize{\bf Fig.~1.} J-V characteristics of the TOPCon silicon solar cell. The comparisons of computer time and memory cost between SolarDesign platform and Silvaco are also presented. 1000 sampling points are adopted for the J-V characteristics.}
\end{center}

\begin{center}
    \includegraphics[width=0.8\linewidth]{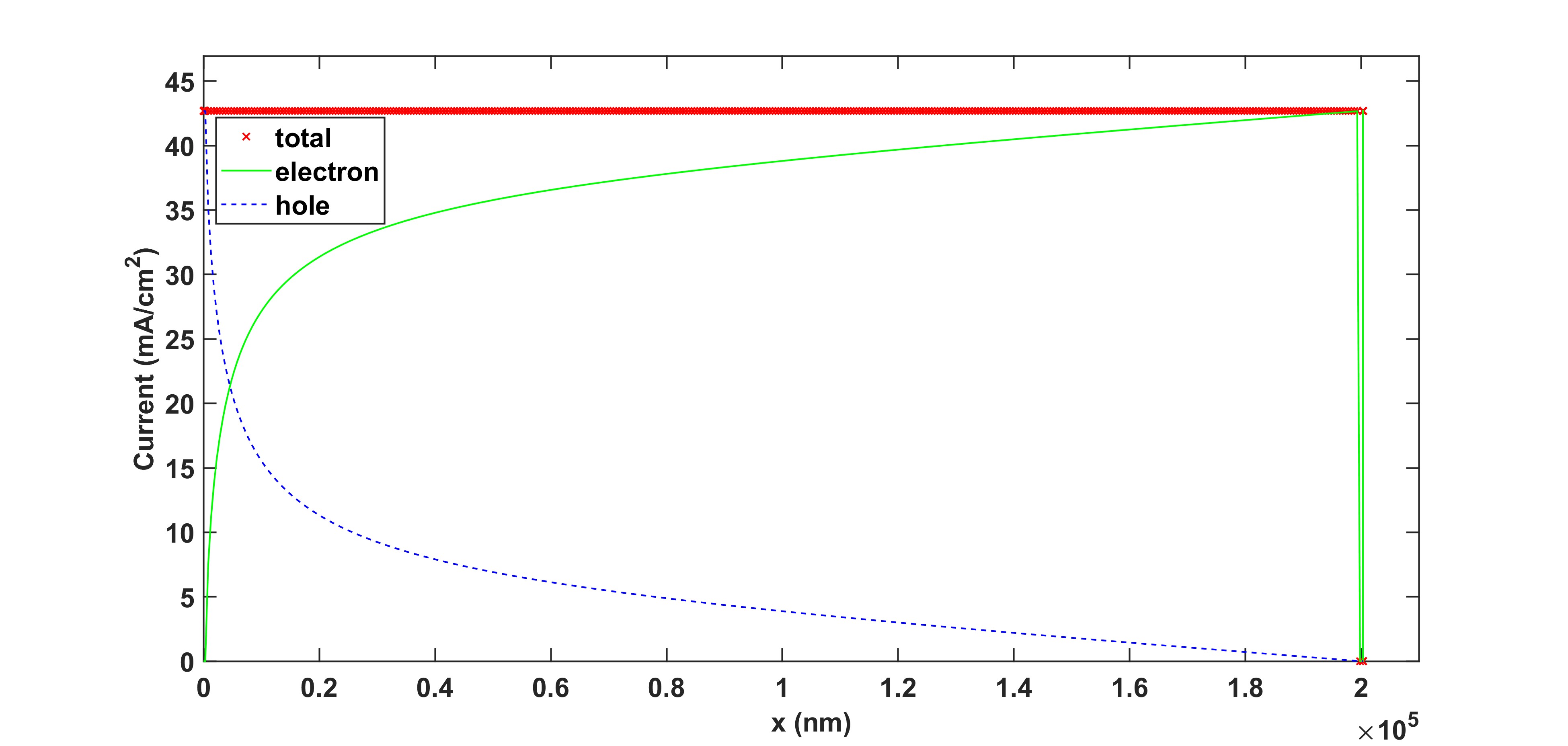}\\[5pt]  
    \parbox[c]{15.0cm}{\footnotesize{\bf Fig.~2.} Classical total, electron, and hole currents of the TOPCon silicon solar cell under the short-circuit condition, where 1 nm silica layer is used for electron current tunneling and surface recombination reduction.}
\end{center}

\subsection{Organic solar cell}
Device structure of an organic solar cell is set as: ITO [70 nm]-ZnO [30 nm]-P3HT:PCBM [200 nm]-PEDOT [50 nm]-Au [100 nm]. Table 3 lists characteristic parameters of the organic cell with different exciton delocalization ratios. As illustrated in Fig. 3, higher Jsc is expected with a larger delocalization ratio of exciton. Moreover, due to reduced built-in potential, exciton dissociation probability decreases when the organic cell is operated from the short-circuit to open-circuit condition (see the inset of Fig. 3).
\begin{table}[h!]
  \begin{center}
    \caption{Characteristic parameters of the organic solar cell with different exciton delocalization ratios.}
    \begin{tabular}{c|c|c|c|c} 
    \hline
      {delocalization ratio} & {Jsc ($\rm{mA/cm^2}$)} & {Voc (V)} & {FF} & {PCE (\%)}\\
      \hline
      0.7 & 12.12 & 0.80 & 0.67 & 6.48 \\
      0.3 & 11.00 & 0.80 & 0.65 & 5.69 \\
      \hline
    \end{tabular}
  \end{center}
\end{table}

\begin{center}
    \includegraphics[width=0.7\linewidth]{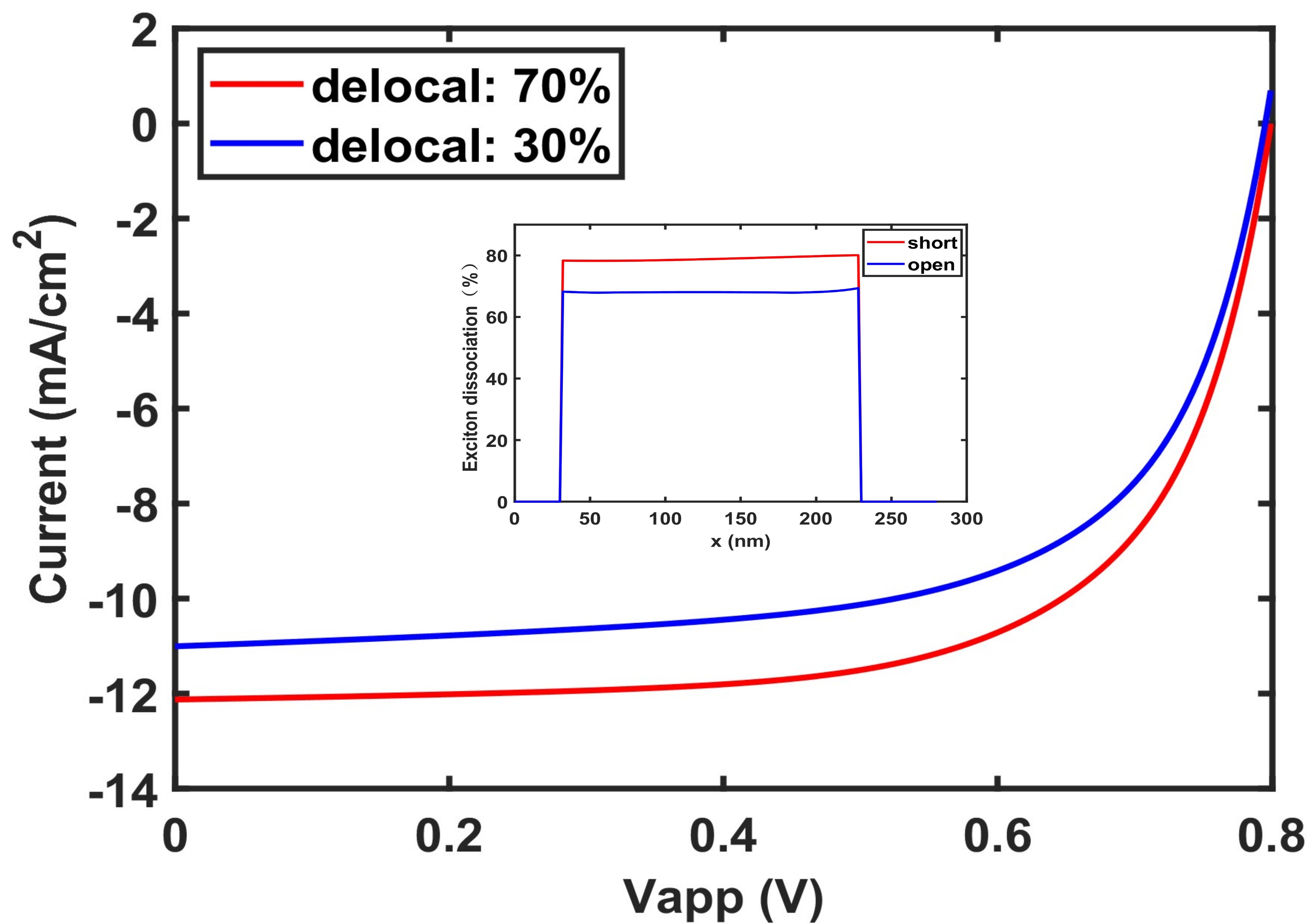}\\[5pt]  
    \parbox[c]{15.0cm}{\footnotesize{\bf Fig.~3.} J-V characteristics of the organic solar cell. “delocal” denotes the delocalization ratio of exciton. The inset shows the exciton dissociation probabilities under the short-circuit and open-circuit conditions.}
\end{center}

\subsection{Perovskite solar cell}
Device structure of a perovskite solar cell is set as: ITO [50 nm]-Ni$\rm{O}_{X}$ [20 nm]-MAPb$\rm{I}_3$ [500 nm]-$\rm{C}_{60}$ [50 nm]-Ag [120 nm]. The Jsc, Voc, FF, and PCE of the perovskite cell are $21.68\,\,\mathrm{mA/cm^2}$, 0.99 V, 0.81, and 17.38\%, respectively. Compared to COMSOL, the calculation error of J-V characteristics by SolarDesign is lower than 1\% (See Fig. 4). Particularly, the computer time is reduced by more than 100 times. Figure 5 shows energy band diagram of the perovskite cell under the open-circuit condition. The quasi-Fermi levels of electron and hole remain constant, which indicates zero currents.  Furthermore, hysteresis behaviors are simulated under a series of voltage scanning rates as depicted in Fig. 6. The J-V characteristics are traced in forward (from short-circuit to open-circuit) and reverse (from open-circuit to short-circuit) scans. The hysteresis effect can be ignored at very fast or very slow scanning rate. The significant difference of the Jsc between the forward and reverse scans indicates dominated bulk recombination at the perovskite active layer. Additionally, the ion migration results by SolarDesign have been benchmarked against those by IonMonger~\cite{3,4}.

\begin{center}
    \includegraphics[width=0.8\linewidth]{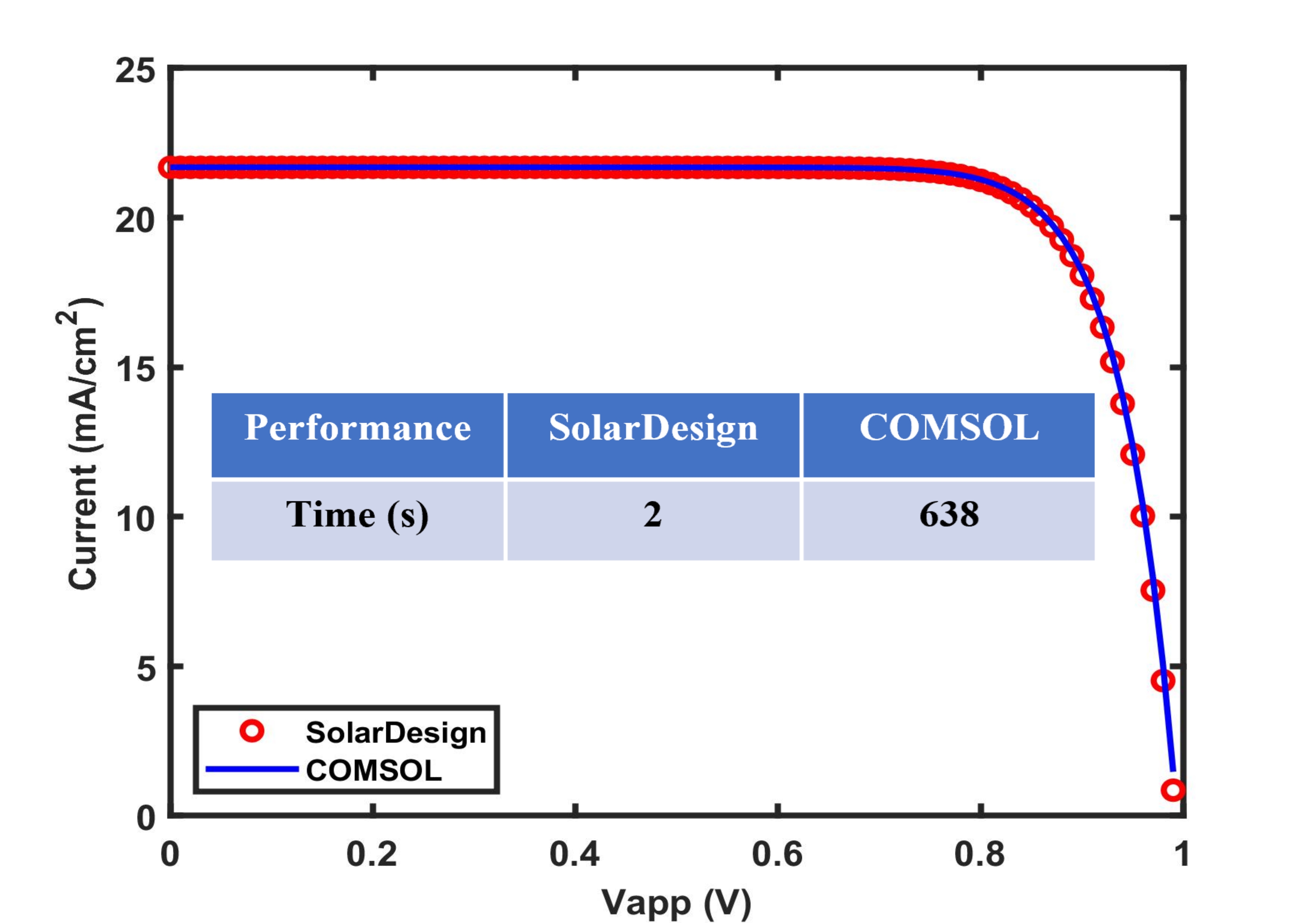}\\[5pt]  
    \parbox[c]{15.0cm}{\footnotesize{\bf Fig.~4.} J-V characteristics of the perovskite solar cell. The comparison of computer time between SolarDesign platform and COMSOL is also presented. 1000 sampling points are adopted for the J-V characteristics.}
\end{center}

\begin{center}
    \includegraphics[width=0.8\linewidth]{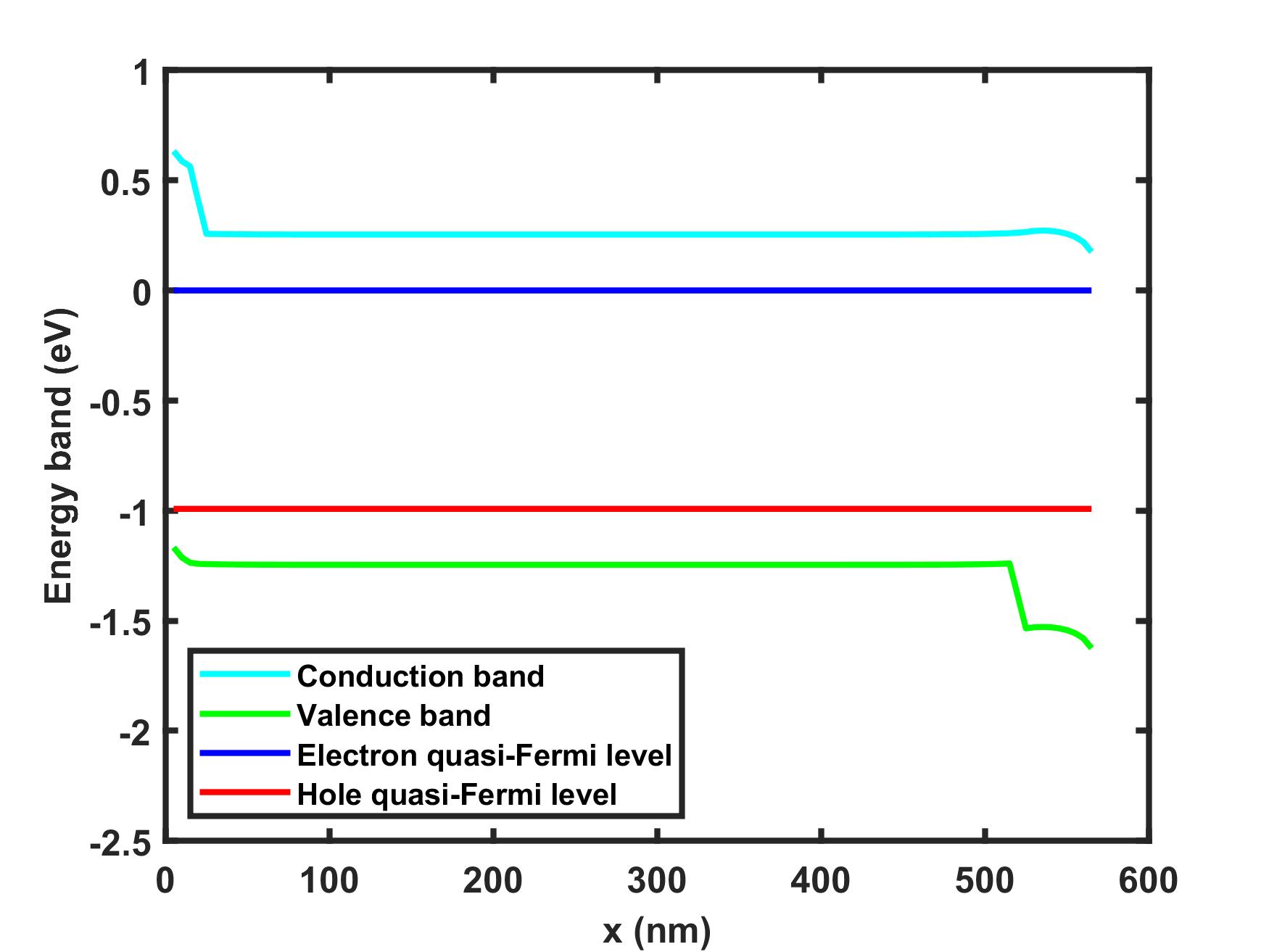}\\[5pt]  
    \parbox[c]{15.0cm}{\footnotesize{\bf Fig.~5.} Energy band diagram of the perovskite solar cell under the open-circuit condition.}
\end{center}

\begin{center}
    \includegraphics[width=0.8\linewidth]{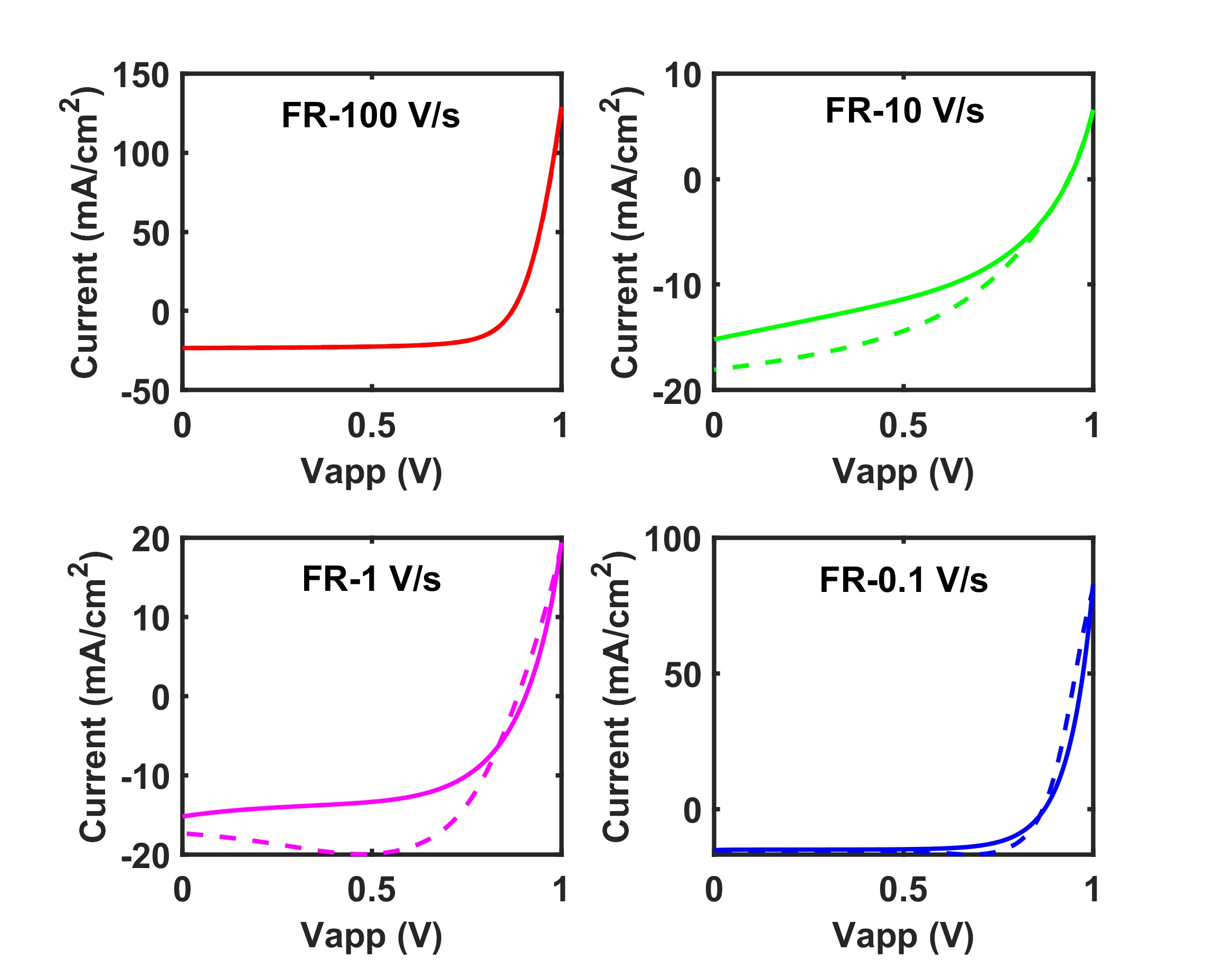}\\[5pt]  
    \parbox[c]{15.0cm}{\footnotesize{\bf Fig.~6.} J-V hysteresis of the perovskite solar cell with different voltage scanning rates (V/s). The forward-reverse (FR) scanning mode is adopted. Solid and dashed lines denote forward and reverse scans, respectively.}
\end{center}

\subsection{Quantifying efficiency losses of solar cells}
The efficiency losses of a perovskite solar cell incorporating star-shaped polymer~\cite{19} are quantified by a modified diode model~\cite{18} in the SolarDesign platform. The averaged fitting errors for the measured J-V characteristics are only $\sim 0.1$ $\rm{mA}/\rm{cm}^2$ for both the control and polymer incorporated cells, where hysteresis effects are ignorable and thus only forward scanning results are analyzed. Figure 7 shows that the incorporation of polymer reduces the bulk recombination and increases both the series and shunt resistances. The dominated surface recombination in the polymer incorporated cell results in a very high FF of 0.86.
    
\begin{center}
    \includegraphics[width=0.8\linewidth]{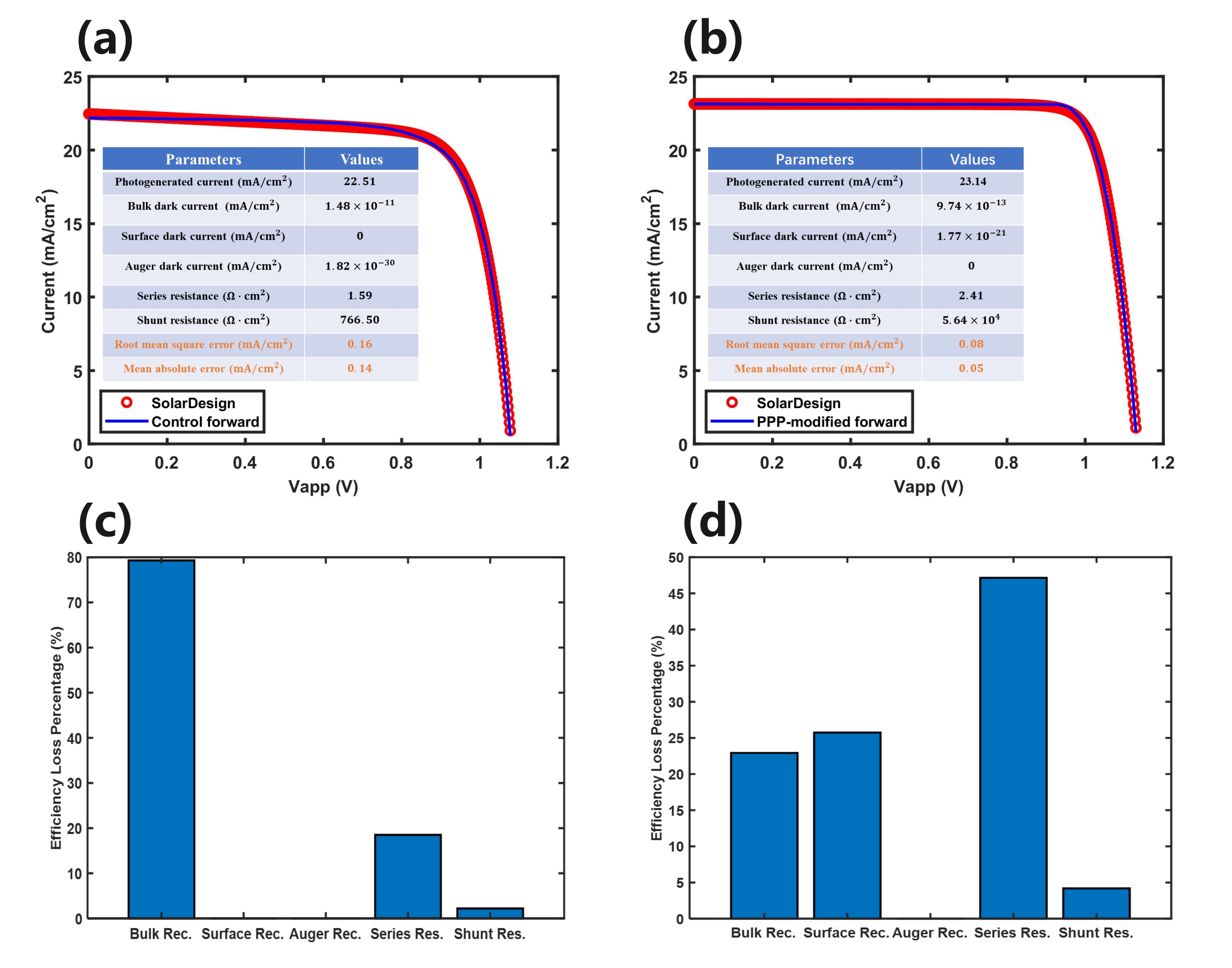}\\[5pt]  
    \parbox[c]{15.0cm}{\footnotesize{\bf Fig.~7.} J-V characteristics and efficiency losses of perovskite solar cells. (a, b) show the J-V characteristics of the control and polymer incorporated (PPP-modified) cells, respectively; the insets show the circuit parameters and fitting errors. (c, d) show the quantified efficiency losses of the control and polymer incorporated cells, respectively.}
\end{center}

\section{Conclusion}

SolarDesign is a comprehensive, powerful, and user-friendly photovoltaic device simulation and design platform. The platform adopts a multi-level, multi-scale simulation architecture, realizing the full process of photovoltaic design from first-principles material simulation to multi-physics device simulation and compact circuit simulation. Compared to mainstream commercial software, the platform's simulation results are reliable, with calculation speeds improved by more than an order of magnitude. It can also model and analyze emerging solar cells such as organic and perovskite devices. The platform currently has over 1,000 registered users from more than 300 organizations including 90 companies.

Future development of the platform will focus on the following aspects: (1) Implementing large-scale photovoltaic module simulation; (2) Realizing photovoltaic testing and characterization simulation (including transient photoluminescence spectroscopy, transient photocurrent/photovoltage spectroscopy, and impedance spectroscopy); (3) Enhancing the photovoltaic material and device libraries to foster an ecosystem in the photovoltaic field.




\addcontentsline{toc}{chapter}{References}
\begin{thebibliography}{99}\footnotesize
\itemsep=-3pt plus.2pt minus.2pt   

\bibitem{1}
Tian L, Sha W E I, Xie H, Liu D, Sun T G, Xia Y S, et al.
\href{https://doi.org/10.1063/5.0209479}
{2024 \emph{Journal of Applied Physics} \textbf{135} 225703}

\bibitem{2}
Wang Z S, Sha W E I, and Choy W C H  
\href{https://doi.org/10.1063/1.4970958}
{2016 \emph{Journal of Applied Physics} \textbf{120} 213101}

\bibitem{3}
Courtier N E, Cave J M, Walker A B, Richardson G, and Foster J M  
\href{https://doi.org/10.1007/s10825-019-01396-2}
{2019 \emph{Journal of Computational Electronics} \textbf{18} 1435-1449}

\bibitem{4}
Courtier N E, Cave J M, Foster J M, Walker A B, and Richardson G
\href{https://doi.org/10.1039/C8EE01576G}
{2019 \emph{Energy $\&$ Environmental Science} \textbf{12} 396-409}

\bibitem{5}
\href{https://www.comsol.com/}{https://www.comsol.com/}

\bibitem{6}
\href{https://silvaco.com/}{https://silvaco.com/}

\bibitem{7}
\href{https://www.ansys.com/products/optics/fdtd}{https://www.ansys.com/products/optics/fdtd}

\bibitem{8}
\href{https://www.pvlighthouse.com.au/}{https://www.pvlighthouse.com.au/}

\bibitem{9}
\href{https://www.pveducation.org/}{https://www.pveducation.org/}

\bibitem{20}
Hohenberg P and Kohn W
\href{https://doi.org/10.1103/PhysRev.136.B864}
{1964 \emph{Phys. Rev.} \textbf{136} B864}

\bibitem{21}
Kresse G and Furthmüller J
\href{https://doi.org/10.1103/PhysRevB.54.11169}
{1996 \emph{Phys. Rev. B} \textbf{54} 11169}

\bibitem{22}
Kresse G and Furthmüller J
\href{https://doi.org/10.1016/0927-0256(96)00008-0}
{1996 \emph{Computational Materials Science} \textbf{6} 15}

\bibitem{23}
Blöchl P E
\href{https://doi.org/10.1103/PhysRevB.50.17953}
{1994 \emph{Phys. Rev. B} \textbf{50} 17953}

\bibitem{24}
Perdew J P, Burke K, and Ernzerhof M
\href{https://doi.org/10.1103/PhysRevLett.77.3865}
{1996 \emph{Phys. Rev. Lett.} \textbf{77} 18}

\bibitem{25}
Paier J, Marsman M, Hummer K, Kresse G, Gerber I C, and Ángyán J G
\href{https://doi.org/10.1063/1.2187006}
{2006 \emph{J. Chem. Phys.} \textbf{124} 154709}

\bibitem{26}
Heyd J, Scuseria G E, and Ernzerhof M
\href{https://doi.org/10.1063/1.1564060}
{2003 \emph{J. Chem. Phys.} \textbf{118} 8207}

\bibitem{27}
Klimeš J, Bowler D R, and Michaelides A
\href{https://doi.org/10.1088/0953-8984/22/2/022201}
{2009 \emph{J. Phys.: Condens. Matter} \textbf{22} 022201}

\bibitem{28}
Zhao X G, et al.
\href{https://doi.org/10.1016/j.scib.2021.06.011}
{2021 \emph{Science Bulletin} \textbf{66} 1973}

\bibitem{29}
Luo S, et al.
\href{https://doi.org/10.1021/acs.jpca.2c03416}
{2022 \emph{J. Phys. Chem. A} \textbf{126} 4300}

\bibitem{30}
Gajdoš M, Hummer K, Kresse G, Furthmüller J, and Bechstedt F
\href{https://doi.org/10.1103/PhysRevB.73.045112}
{2006 \emph{Phys. Rev. B} \textbf{73} 045112}

\bibitem{31}
Tanaka K, Takahashi T, Ban T, Kondo T, Uchida K, and Miura N
\href{https://doi.org/10.1016/S0038-1098(03)00566-0}
{2003 \emph{Solid State Communications} \textbf{127} 619}

\bibitem{10}
Chew W C 1984 \emph{Waves and Fields in Inhomogenous Media} (New York: Wiley-IEEE Press) 

\bibitem{11}
Sha W E I, Choy W C H, Liu Y G, and Chew W C
\href{https://doi.org/10.1063/1.3638466}
{2011 \emph{Applied Physics Letters} \textbf{99} 113304}

\bibitem{12}
Sha W E I, Ren X, Chen L, and Choy W C H
\href{https://doi.org/10.1063/1.4922150}
{2015 \emph{Applied Physics Letters} \textbf{106} 221104}

\bibitem{13}
Ren X, Wang Z, Sha W E I, and Choy W C H
\href{https://doi.org/10.1021/acsphotonics.6b01043}
{2017 \emph{ACS Photonics} \textbf{4} 934-942}

\bibitem{14}
Selberherr S 1984 \emph{Analysis and Simulation of Semiconductor Devices} (New York: Springer) 

\bibitem{15}
Sha W E I, Choy W C H, Wu Y, and Chew W C
\href{https://doi.org/10.1364/OE.20.002572}
{2012 \emph{Optics Express} \textbf{20} 2572-2580}

\bibitem{16}
Sha W E I, Zhang H, Wang Z S, Zhu H L, Ren X, Lin F, et al.
\href{https://doi.org/10.1002/aenm.201701586}
{2018 \emph{Advanced Energy Materials} \textbf{8} 1701586}

\bibitem{17}
Xu T, Wang Z S, Li X H, and Sha W E I
\href{https://doi.org/10.7498/aps.70.20201975}
{2021 \emph{Acta Physica Sinica} \textbf{70} 098801}

\bibitem{18}
Lin M, Xu X, Tian H, Yang Y, Sha W E I, and Zhong W
\href{https://doi.org/10.1002/solr.202300722}
{2024 \emph{Solar RRL} \textbf{8} 2300722}

\bibitem{19}
Cao Q, Li Y, Zhang H, Yang J, Han J, Xu T, et al.
\href{https://doi.org/10.1126/sciadv.abg0633}
{2021 \emph{Science Advances} \textbf{7} eabg0633}





\end{thebibliography}
\end{document}